\documentclass{ws-procs975x65}

\begin{document}

\title{Generalized Parton Distributions and Nucleon Resonances}

\author{M. Guidal, S. Bouchigny, J.-P. Didelez, C. Hadjidakis, E. Hourany}

\address{Institut de Physique Nucl\'eaire Orsay, F-91406 Orsay, France}

\author{M. Vanderhaeghen}

\address{Institut f\"ur Kernphysik, Johannes Gutenberg-Universit\"at, 
        D-55099 Mainz, Germany}  


\maketitle

\abstracts{We discuss the relations between 
Generalized Parton Distributions (GPDs) and nucleon resonances. We 
first briefly introduce the concept of ``transition" GPDs. Then we 
discuss a straightforward application to the modelization of the N-$\Delta$ 
magnetic transition form factor. Finally, we discuss the experimental aspects 
of the subject and present first preliminary experimental investigations
in this field.}

\section{Brief introduction to ``transition" GPDs}

The Generalized Parton Distributions provide one of the most 
complete information on the structure of the nucleon~:
they allow to access the longitudinal momentum distributions 
of the partons in the nucleon as well as the nucleon's transverse profile, 
the parton correlations, their orbital momentum contribution to 
the total nucleon's spin, quark-antiquark components in the nucleon, etc...

The GPDs can be accessed through the hard (i.e. large photon 
virtuality $Q^2$) exclusive electroproduction of photons (``Deep 
Virtual Compton Scattering" -DVCS-) and mesons 
-$\pi^{0,\pm}$, $\rho^{0,\pm}$, $\omega$, $\phi$, etc...- on the nucleon.

The formalism introducing the Generalized Parton Distributions 
can be found in Refs.~\cite{muller,Ji97,Rady,Collins97} and 
is summarized in the 
contribution of A. Belitstky to these proceedings. We refer the reader 
to these articles for the standard definitions and notations used 
in the following. 

The GPD formalism was originally developed for exclusive reactions 
on the nucleon, both in the initial and final states, but was recently 
extended to more general baryonic final states~\cite{prl84}, 
in particular for the simplest non-elastic transition 
the N-$\Delta$ case. Like ``transition" N-$\Delta$ form factors, one can
introduce about ``transition" N-$\Delta$ GPDs and these are in principle 
different from the nucleon GPDs.

N-$\Delta$ GPDs contain the same type of information as the
nucleon GPDs, i.e. the quarks longitudinal momentum distributions 
and their transverse positions in the $\Delta$, their spin distribution, 
etc... and, in the case of $\Delta$-$\Delta$ GPDs, similarly to Ji's sum rule, 
the orbital contribution of the quarks to the spin of the $\Delta$.

At leading twist in QCD, the N-$\Delta$ transition can be parametrized
in terms of 3 vector N-$\Delta$ GPDs and 4 axial-vector N-$\Delta$ GPDs.
Among them, one expects 3 N-$\Delta$ GPDs to dominate at small $\mid t\mid$~:
the magnetic vector N-$\Delta$ GPD $H_M$, whose first moment corresponds
to the N-$\Delta$ magnetic transition form factor $G_M^*(t)$, and the
axial vector N-$\Delta$ GPDs $C_1$ and $C_2$ whose first moments
correspond to the axial and pseudoscalar N-$\Delta$ form factors
respectively.

\begin{figure}[ht]
\epsfxsize=15 cm
\epsfysize=10.cm
\centerline{\epsffile{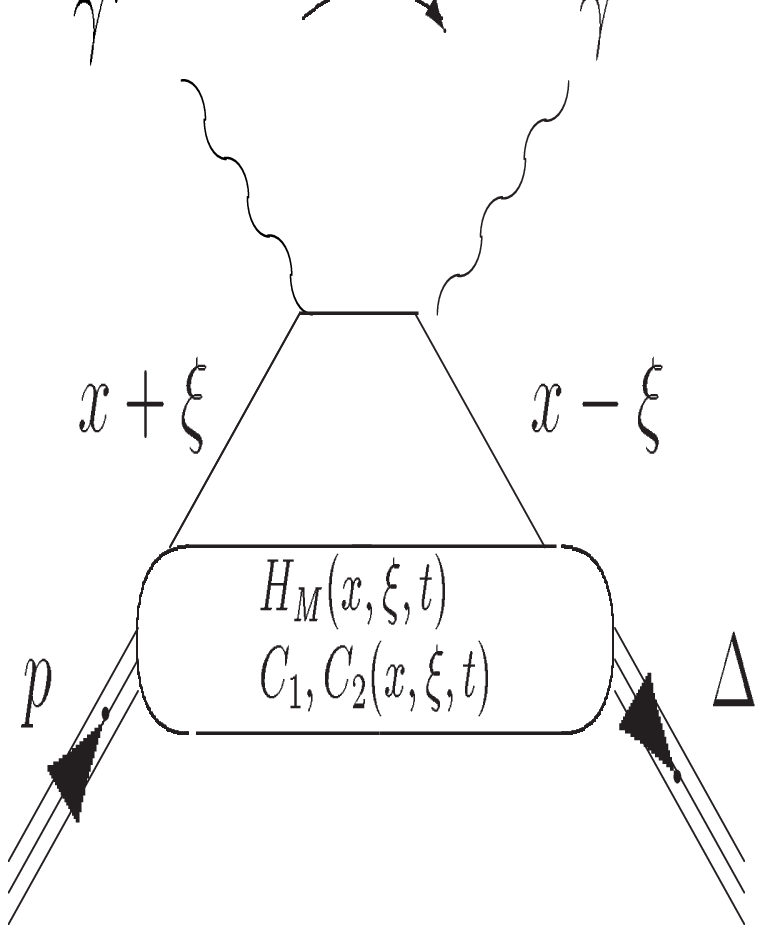}}
\vspace{-5cm}
\caption[]{The exclusive process $ep\to e^\prime\Delta\gamma$
expressed in terms of the 3 $N-\Delta$ transition GPDs, which are 
expected to dominate at small $\mid t\mid$~: $C_1, C_2$ and $H_M$.
The longitudinal momentum fractions of the quarks $(x,\xi)$ are 
indicated on the figure as well as the definition of the 
momentum transfer $t$.}
\label{diag}
\end{figure}

In order to provide numerical estimates for the N-$\Delta$ amplitudes,
these three GPDs have been estimated in Ref.~\cite{prl84} in the large
$N_c$ limit. In this limit (valid to about 30\%), they can 
be related to three nucleon GPDs~: $C_1 \propto \tilde H^{(3)}, 
C_2 \propto \tilde E^{(3)}$ and $H_M \propto E^{(3)}$, where the superscript $^{(3)}$
refers to the isovector nucleon GPDs combination. As various modelizations
of the nucleon GPDs are available in the literature, these relations allow 
to make numerical estimates and calculations for observables (cross sections, 
asymmetries,...) related to exclusive reactions involving the N-$\Delta$ 
transition, such as $ep\to e^\prime\Delta\gamma$ (usually refered to
as $\Delta$VCS by analogy with DVCS -see Fig.~\ref{diag}-), 
$ep\to e^\prime\Delta^{++}\pi^-$, etc...

\begin{figure}[ht]
\epsfxsize=15.cm
\epsfysize=10.cm
\centerline{\epsffile{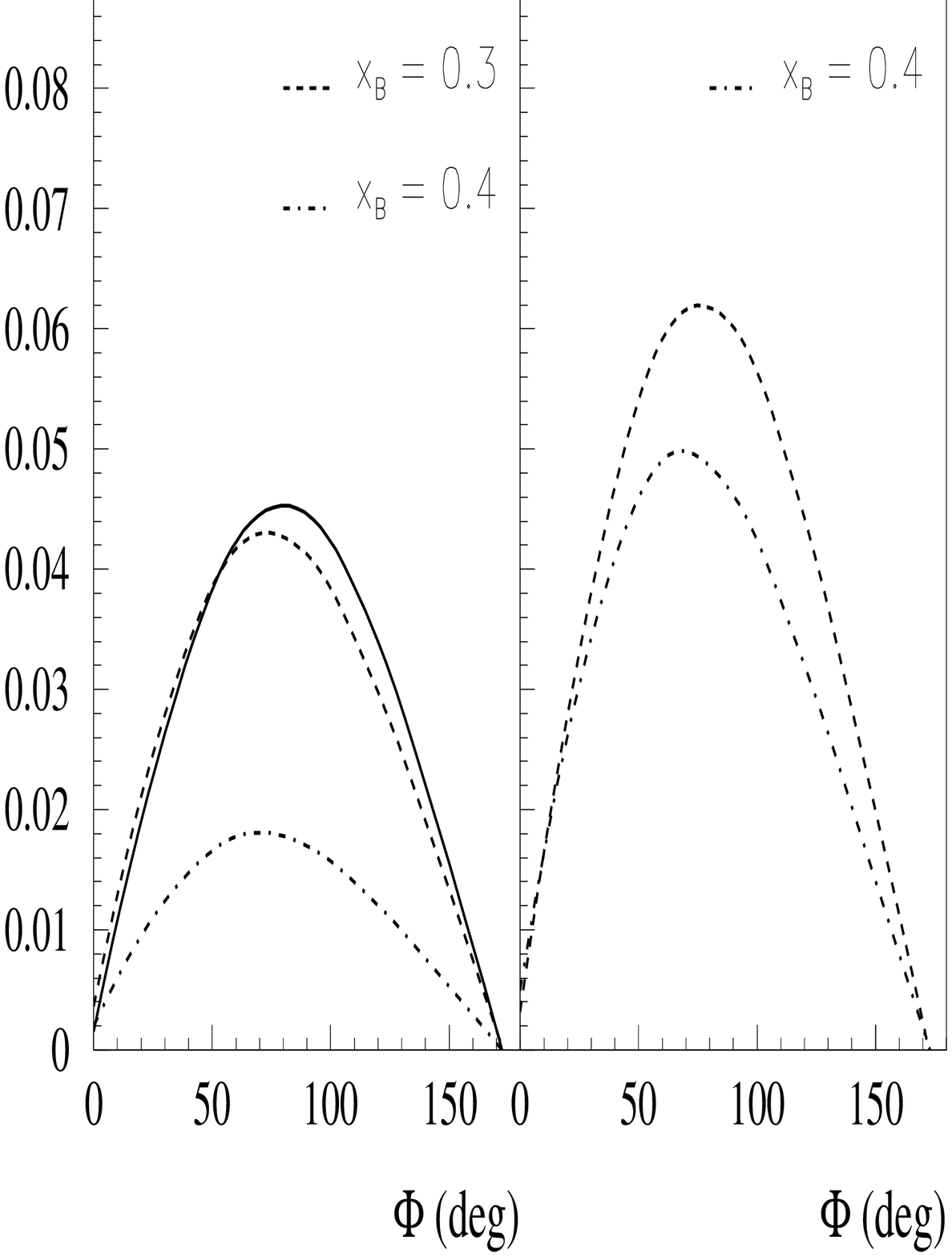}}
\vspace{-2.5cm}
\caption[]{Predictions for the beam helicity asymmetry arising from the
interference of the $\Delta$VCS and the associated Bethe-Heitler 
process for typical JLab kinematics.}
\label{asym-Delta}
\end{figure}

In the case of the $\Delta$VCS process, similarly to the 
DVCS process, there is an interference with the associated Bethe-Heitler 
mechanism (where the outgoing photon is emitted by the incoming or 
scattered electron), and this produces a beam helicity asymmetry, which is 
shown on Fig.~\ref{asym-Delta}.

The large $N_c$ relations, although not extremely precise, allow also to interpret 
the $\Delta$VCS reaction in terms of nucleon GPDs and therefore permits 
to access a different {\it flavor} combination of the nucleon GPDs~: for instance, 
in DVCS, one accesses in general the combination $\frac{4}{9}\tilde H^u
+\frac{1}{9}\tilde H^d$ whereas, in $\Delta$VCS, one
accesses the isovector part $\tilde H^u -\tilde H^d$.
In this way, the $\Delta$VCS process is therefore also useful to
carry out a {\it flavor} decomposition of the nucleon GPDs.

A new way to study baryon spectroscopy is therefore opening up
through these transition $N\to N^*,\Delta^*$ GPDs, where not only the $t$-dependence,
which can currently be determined through transition form factors and
which reflects the transverse spatial distribution of the partons
in the nucleon, can be accessed but also their $x$-dependence, 
reflecting the longitudinal momentum distributions of the quarks in the 
$N^*$'s. 

\section{Modelization of the N-$\Delta$ transition form factor}

In this part, as a basic illustration of the information
contained in the GPDs, we show how the N-$\Delta$ transition form factor
can be estimated from the (nucleon) GPDs, relying on very simple
assumptions. We start from the model independent sum rules which relate the
GPDs to the form factors~:

\begin{eqnarray}
F_{2}^{q}(t) \,=\, \int _{-1}^{+1}dx\; E^{q}(x,\xi ,t) \,  ,
\hspace{2cm} 
G_{M}^{*}(t) \,=\, \int _{-1}^{+1}dx\; H_{M}(x,\xi ,t) \, \; 
\label{eq:hesumrule} 
\end{eqnarray}

where $F_2$ is the nucleon Pauli form factor and $G_M^*$ the N-$\Delta$
magnetic transition form factor.

In the large $N_c$ limit, the N-$\Delta$ GPDs can be expressed 
in terms of the isovector combination of nucleon GPDs and one arrives at 
the following sum rule~\cite{prl84}~: 
  
\begin{eqnarray}
G_M^{*}(t) \,=\, {{G_M^{*}(0)} \over {\kappa_V}} \; \int _{-1}^{+1}dx\; 
\biggl \{ E^{u}(x,\xi ,t) \,-\, E^{d}(x,\xi ,t) \biggr \} 
\; =\;  {{G_M^{*}(0)} \over {\kappa_V}} \; 
\biggl \{ F_2^p(t) - F_2^n(t) \biggr \} \, 
\label{eq:gmsumrule} 
\end{eqnarray}
where $\kappa_V = \kappa_p - \kappa_n = 3.70$.
In the large $N_c$ limit, the value $G_M^{*}(0)$ is given by~\cite{GPV01}:
$G_M^{*}(0) = \kappa_V / \sqrt{3}$, some 30 \% smaller than the
experimental number. In calculations, we therefore use the
phenomenological value $G_M^{*}(0) \approx 3.02$~\cite{Tiator00}.

\begin{figure}[ht]
\epsfxsize=15 cm
\epsfysize=10.cm
\centerline{\epsffile{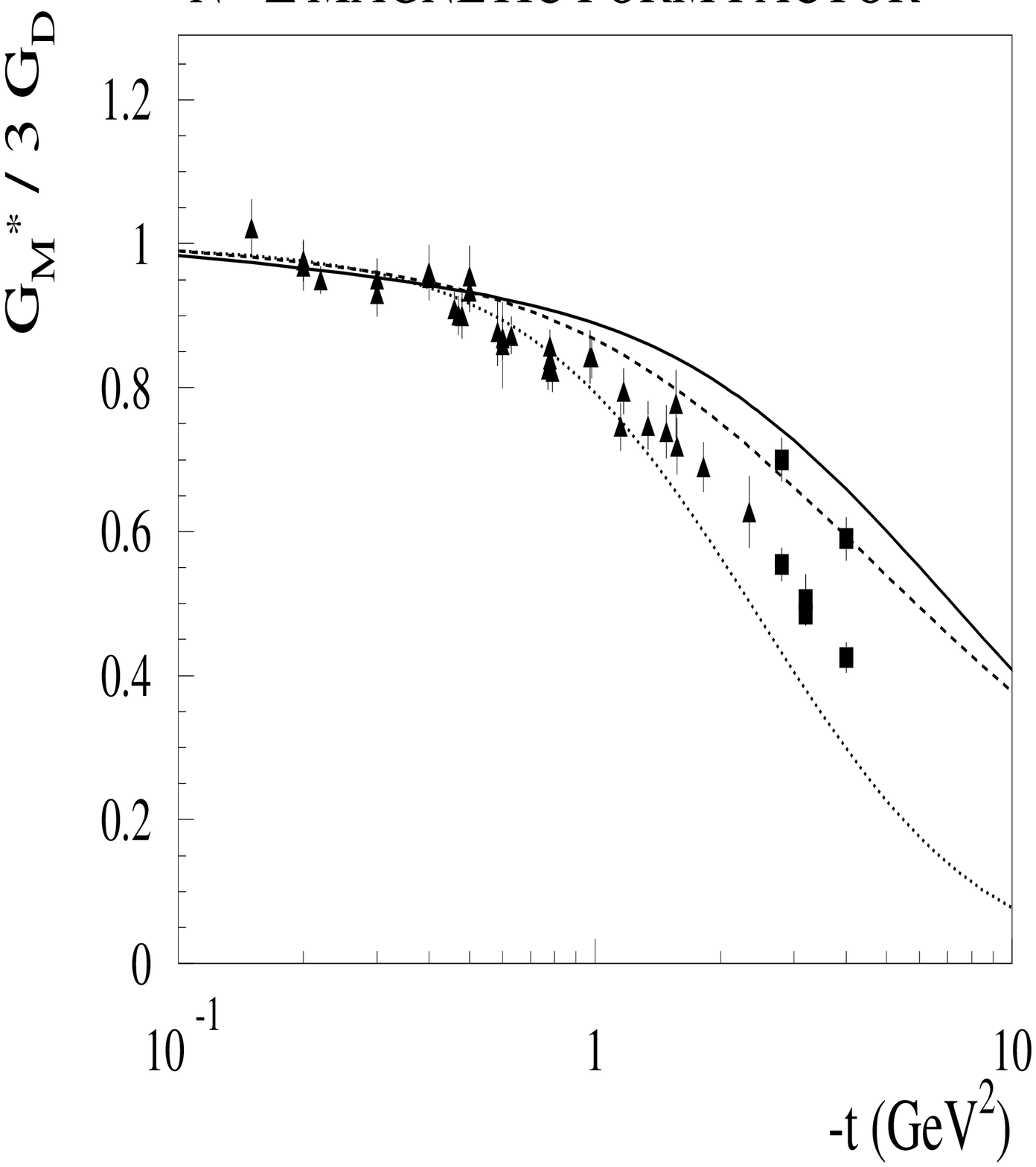}}
%\vspace{-5cm}
\vspace{-2.5cm}
\caption[]{The $N \to \Delta$ magnetic transition form factor, relative 
to the dipole. 
The dotted curve corresponds to $\alpha^{'}$ = 1.102 GeV$^{-2}$. 
The other curves correspond to extensions of the present model 
by allowing for non-linear Regge trajectories (see Ref.~\cite{exclusive}
for a more detailed discussion). 
The data for $G_M^{*}$ are from the compilation of
Ref.~\cite{Tiator00}. For the JLab data points at 2.8 and 4 GeV$^2$, 
both the analyses of \cite{Frolov99} (upper points) and
\cite{Tiator00} (lower points) are shown.}
\label{gmdel}
\end{figure}

One can choose $\xi = 0$ in the previous equations, and by  
modeling the nucleon GPD $E(x,0,t)$, one can therefore estimate the
N-$\Delta$ transition form factor. A plausible ansatz at low $\mid t\mid$, 
inspired from a Regge form, for the GPD $E(x, 0, t)$ could be~\cite{GPV01}:
\begin{eqnarray}
E^q(x,0,t) \,&=&\, \kappa_q q_v(x) \;{{1} \over {x^{\alpha^{\, '} t}}} \; ,
\label{eq:f2d_1}
\end{eqnarray}
where $q_v(x)$ is the (u,d) valence quark distribution normalized to (2,1)
respectively and where $\kappa_u$ and $\kappa_d$ are given by 
\begin{eqnarray}
\kappa_u \,&=&\, 2 \, \kappa_p \;+\; \kappa_n \; , 
\label{eq:kappau} \\
\kappa_d \,&=&\, \kappa_p \;+\; 2 \, \kappa_n \; .
\label{eq:kappad}
\end{eqnarray}
In equation~(\ref{eq:f2d_1}), $\alpha'(t)$ is the slope of standard Regge 
trajectory (to be fitted)
around 1 GeV$^2$. It has been shown in Ref.~\cite{ferrara} that such kind of 
ansatz was able to reproduce rather well the electric and magnetic radii of the
nucleon and the Dirac and Pauli form factors at low $\mid t\mid$ ($<$ 1 GeV$^2$). 
We show on figure~\ref{gmdel} the decent agreement for the N-$\Delta$ transition 
form factor.

\section{Experimental aspects}

\begin{figure}[ht]
\epsfxsize=15 cm
\epsfysize=10.cm
\centerline{\epsffile{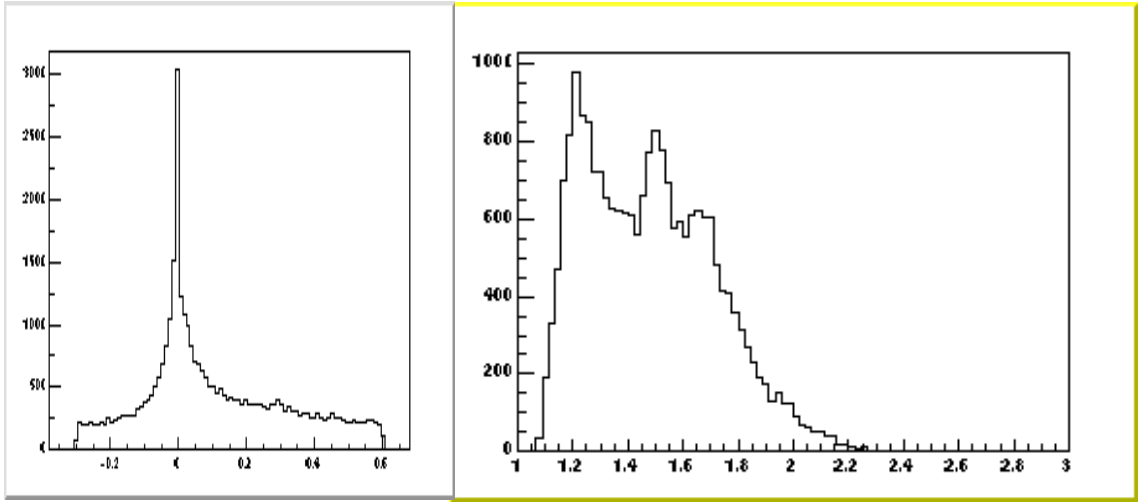}}
\vspace{-2cm}
\caption[]{a (Left)~: Squared missing mass spectrum 
$e^\prime n\pi^+ X$ in GeV$^2$, corresponding to the missing mass of
a $\gamma$ or a $\pi^0$. b (Right)~: Invariant 
mass spectrum of the $n\pi^+$ system in GeV for $W>2$ GeV, therefore corresponding to $N^*(\gamma,\pi^0)$ final states.}
\label{spectra}
\end{figure}

Finally, on the experimental side, first hints of the observation of 
the $ep\to e^\prime\Delta\gamma$ reaction in the CLAS detector at JLab
are showing up. The analysis, led by S. Bouchigny from IPN Orsay, 
has consisted in selecting $e^\prime n \pi^+$ final 
states in data taken with a 4.2 GeV incident electron beam. 
Selecting events whose missing mass MM($e^\prime n \pi^+ X$) is around zero
\footnote{It should be kept in mind that there is
certainly a contamination of the $\Delta$VCS signal by  
$\Delta\pi^0$ final states at this preliminary stage of analysis, the 
current resolution of CLAS not permitting the separation of the two by
the missing mass technique.},
in order to be compatible with a missing photon (see figure~\ref{spectra}-a), and 
imposing a $W>2$ GeV cut so that one is above the resonance region, one was 
able to observe the $n\pi^+$ invariant mass spectra of 
figure~\ref{spectra}-b where one clearly distinguishes, not 
only $\Delta\gamma$ final states, but also higher $N^*\gamma$ final states,
providing evidence for N$\to$N$^*$ DVCS processes.

%\begin{figure}[ht]
%\epsfxsize=15 cm
%\epsfysize=10.cm
%\centerline{\epsffile{mm_enpiX.eps}}
%\centerline{\epsffile{im_npi.eps}}
%\vspace{-13.6cm}
%\caption[]{a (Left)~: Squared missing mass spectrum 
%$e^\prime n\pi^+ X$ in GeV$^2$, corresponding to the missing mass of
%a $\gamma$ or a $\pi^0$. The peak is not precisely centered at 0
%because b (Right)~: Invariant 
%mass spectrum of the $n\pi^+$ system in GeV for $W>2$ GeV, therefore corresponding to $N^*(\gamma,\pi^0)$ final states.}
%\label{spectra}
%\end{figure}

The statistics in this figure (effectively equivalent to a few days of beam
time) are not enough yet to extract a significant beam asymmetry to be 
compared to the theoretical calculations of figure~\ref{asym-Delta} but the observed 
signals are certainly very encouraging.

Data have been taken last year in CLAS with a 5.75 GeV beam and are in the process
of being analysed (with 5 times more statistics than with the 4.2 GeV run of 
figure~\ref{spectra}) and should allow to have a first glance at a
$\Delta$VCS beam asymmetry very soon.

Finally, the forthcoming equipment of CLAS with a fine (angular and energy) resolution 
and high counting rate capability calorimeter, to be installed and operational 
mid-2004 (see for instance Ref.~\cite{e01113}), will also boost tremendously the potential of analysis of 
this channel. The full detection of the photon(s) of the final state will be
possible and will unambiguously
allow to distinguish $e^\prime\Delta\gamma$ from $e^\prime\Delta\pi^0$ final
states.

\section{Conclusion}

We have briefly discussed here a couple of aspects of the relation between 
(transition) Generalized Parton Distributions and nucleon resonances. 
The exclusive electroproduction of a photon or of a meson with baryonic
resonances opens a brand new way to understand the nucleon excitation spectrum,
allowing to access a wealth of information not contained in direct (i.e.
$s$-channel) resonance excitation. This opens up the perspective to
arrive at a 3-dimensional picture of the nucleon and 
its resonances through the formalism of the GPDs.

\end{document}